\documentclass[twocolumn,preprintnumbers,amsmath,amssymb]{revtex4}

\usepackage{graphicx}
\usepackage{dcolumn}
\usepackage{bm}
\usepackage{slashbox}

\begin{document}

\preprint{}

\title{Absence of a Spin Liquid Phase in the Hubbard Model on the Honeycomb Lattice}
\author{Sandro Sorella$^{1,2,3}$}
 \email{sorella@sissa.it}
\author{Yuichi Otsuka$^{3}$}
 \email{otsukay@riken.jp}
\author{Seiji Yunoki$^{3,4,5}$}
 \email{yunoki@riken.jp}
\affiliation{%
$^1$SISSA -- International School for Advanced Studies, Via Bonomea 265, 34136 Trieste, Italy, \\
$^2$Democritos Simulation Center, CNR -- IOM Instituto Officina dei Materiali, 34151 Trieste, Italy, \\
$^3$Computational Materials Science Research Team, RIKEN AICS, Kobe, Hyogo 650-0047, Japan, \\
$^4$Computational Condensed Matter Physics Laboratory, RIKEN ASI, Saitama 351-0198, Japan, and \\
$^5$CREST, Japan Science and Technology (JST), Kawaguchi, Saitama 332-0012, Japan
}%

\date{\today}

\begin{abstract}
A spin liquid is a novel quantum state of matter with no conventional order parameter where a finite charge gap exists
even though the band theory would 
predict metallic behavior. Finding a stable spin liquid in two or higher spatial dimensions is one of the most challenging and 
debated issues in condensed matter physics. Very recently, it has been reported that a model of graphene, i.e., the Hubbard 
model on the honeycomb lattice, can show a spin liquid ground state in a wide region of the phase diagram, between a 
semi-metal (SM) and an antiferromagnetic insulator (AFMI). 
Here, by performing numerically exact quantum Monte Carlo simulations,
we extend the previous study to much larger clusters (containing up to 2592 sites),
and find, if any, a very weak evidence of this spin liquid region. 
Instead, our calculations strongly indicate a direct and continuous quantum phase transition between SM and AFMI. 

\end{abstract}

\maketitle

\section*{Introduction}
A spin liquid can be considered as a Mott insulator that is not adiabatically connected to any band insulator and does not break any symmetry even at zero temperature. 
Recently, much attention 
has been focused on a possible spin liquid in two or three spatial dimensions~\cite{balents2,muramatsu,white,balents,kivelson}. 
On one hand, it has been suggested experimentally that several organic materials represent  good candidates for spin 
liquids~\cite{organic1,organic2,organic3}. 
On the other hand, 
the existence of a spin liquid has so far been demonstrated only in very few and particularly simplified models~\cite{triang,kitaev}.

In order to understand this issue, let us consider a model Hamiltonian on a lattice describing the insulating state of electrons at 
half-filling, i.e., one electron per lattice site. Since the charge gap is assumed, 
only spin degrees of freedom remain and can be described by the spin-1/2 Heisenberg model. 
Since any spin-1/2 model corresponds to a well defined interacting hard core boson model, the crucial question is 
to understand how -- at zero temperature -- these bosonic degrees of freedom can avoid Bose-Einstein condensation and/or crystallization, necessary conditions  for a stable spin liquid with no long-range order of any kind.

\begin{figure}
\begin{center}
\includegraphics[width=0.75\hsize]{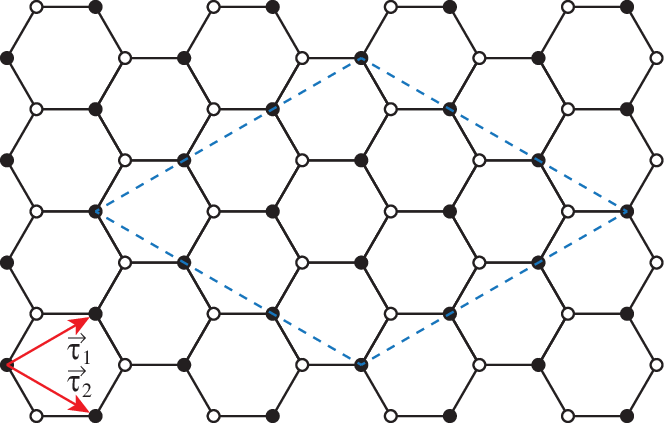}
\caption{\label{lattice}
{\bf The honeycomb lattice.} 
Primitive lattice vectors ${\vec \tau}_1$ and ${\vec \tau}_2$ are denoted by red arrows. As an example, 
the honeycomb lattice with $L=3$ is indicated by dashed blue lines. Solid and open circles indicate sites 
on A and B sublattices, respectively.  
 }
\end{center}
\end{figure}

In this report, we study the ground state of the half-filled Hubbard model on the honeycomb 
lattice (see Fig.~\ref{lattice}) defined by the following Hamiltonian, 
\begin{equation}
{\hat H}=-t\sum_{\langle i,j\rangle,\sigma} \left( c^\dag_{i\sigma}c_{j\sigma}+c^\dag_{j\sigma}c_{i\sigma}\right) + U\sum_i n_{i\uparrow} n_{i\downarrow},
\label{ham}
\end{equation}
where $c^\dag_{i\sigma}$  ($c_{i\sigma}$) is a creation (annihilation) operator of spin up/down ($\sigma=\uparrow,\downarrow$) 
electrons on lattice site $i$, $n_{i\sigma}=c^\dag_{i\sigma}c_{i\sigma}$, and $t$ ($U$) denotes the nearest-neighbor 
hopping (on-site repulsion). This is known to be a model Hamiltonian for graphene with $U/t\approx3$~\cite{neto}. 
More importantly, it is not 
geometrically frustrated, namely, as seen in Fig.~\ref{lattice}, the neighboring sites of any site on A sublattice belong
to B sublattice (and vice versa). Indeed, it is well known that 
the ground state becomes an antiferromagnetic insulator (AFMI), i.e., classically N\'eel ordered, from 
a semi-metal (SM) with increasing $U/t$~\cite{sorella_honey}. 

Very recently, Meng {\it et al.}~\cite{muramatsu} has reexamined the ground state phase diagram of this model and 
found a possible spin liquid phase in the range $3.4 \lessapprox U/t  \lessapprox 4.3$ between SM and AFMI. 
Their finding is rather surprising because it is widely believed 
that a stable spin liquid occurs most likely in frustrated quantum systems where strong quantum fluctuations destroy the long-range
magnetic order even at zero temperature~\cite{balents2}. 
Their study was particularly successful because, with the auxiliary field technique~\cite{hirsch}, there is no sign problem in the
corresponding quantum Monte Carlo simulations, and an accurate finite size scaling was possible by using numerically exact results 
for clusters containing up to  $648$ sites. 
So far, their results represent the most important numerical evidence for a possible spin liquid ground state in a "realistic" 
electronic model in two dimensions (2D), because, to our knowledge, only a particularly simplified quantum dimer model on the triangular lattice~\cite{triang} 
and the Kitaev model~\cite{kitaev}, built {\it ad hoc} to have an exact solution, allow a  spin liquid ground state in 2D. Furthermore, 
their results were considered to be a clear violation of the "Murphy's Law": in a too simple model, not vexed by  the  ''fermion sign problem'', 
nothing interesting can occur~\cite{kivelson}.

Here, by performing simulations for much larger clusters containing up to 2592 sites, we show that antiferromagnetic order 
concomitantly occurs once the insulating behavior sets in, supporting the more conventional Hartree-Fock (HF)  transition from SM to 
AFMI~\cite{sorella_honey}. Although our results agree with the previous study for the same clusters up to 648 sites, we have 
reached a quite different conclusion, as 
the possible spin liquid region reduces substantially to a small interval 
$3.8t\lessapprox U/t \lessapprox 3.9t$, if it ever exists. 
This reminds us similar claims on spin liquid behaviors in different systems in 2D~\cite{santoro,parola}, 
which have been corrected later on by much larger cluster simulations, showing instead antiferromagnetic long-range order~\cite{troyer,sandvik}.

\section*{Results}

We use finite size clusters of $N=2 L^2$ sites (thus containing $L\times L$ unit cells) with periodic boundary conditions 
(see Fig.~\ref{lattice}), which 
satisfy all symmetries of the infinite lattice~\cite{bernu} (also see Supplementary information). Here $L$ is the linear dimension 
of the cluster and we take $L$ up to 36.  
We use the well established auxiliary field Monte Carlo technique~\cite{hirsch}, which allows the statistical evaluation of  
the following quantity,   
\begin{equation}
O(\tau)={ \langle \psi_L |  {\rm e}^{-{\frac{\tau}{2}} {\hat H} } {\hat O} {\rm e}^{ - {\frac{\tau}{2}} {\hat H} } | \psi_R \rangle 
\over  \langle \psi_L | {\rm e}^{ - \tau {\hat H} } | \psi_R \rangle },  
\label{ave}
\end{equation}
where ${\hat O}$ is a physical operator, $|\psi_R \rangle $  ($|\psi_L\rangle $) is the right (left) trial wave function (not orthogonal to the 
exact ground state), and $\tau$ is the projection time. 
The exact ground state expectation value $\langle{\hat O}\rangle$ of the operator ${\hat O}$ is then 
obtained by adopting the limit of $\tau\to \infty$ and $\Delta\tau \to 0$ 
for $O(\tau)$, where $\Delta\tau$ is the  short time discretization of $\tau$. 
This approximation -- the so called Trotter approximation --  is necessary 
to introduce the auxiliary fields~\cite{hirsch} and 
implies a systematic error, negligible for small $\Delta \tau$ (see Supplementary information).

\begin{figure*}
\begin{center}
\includegraphics[width=0.9\hsize]{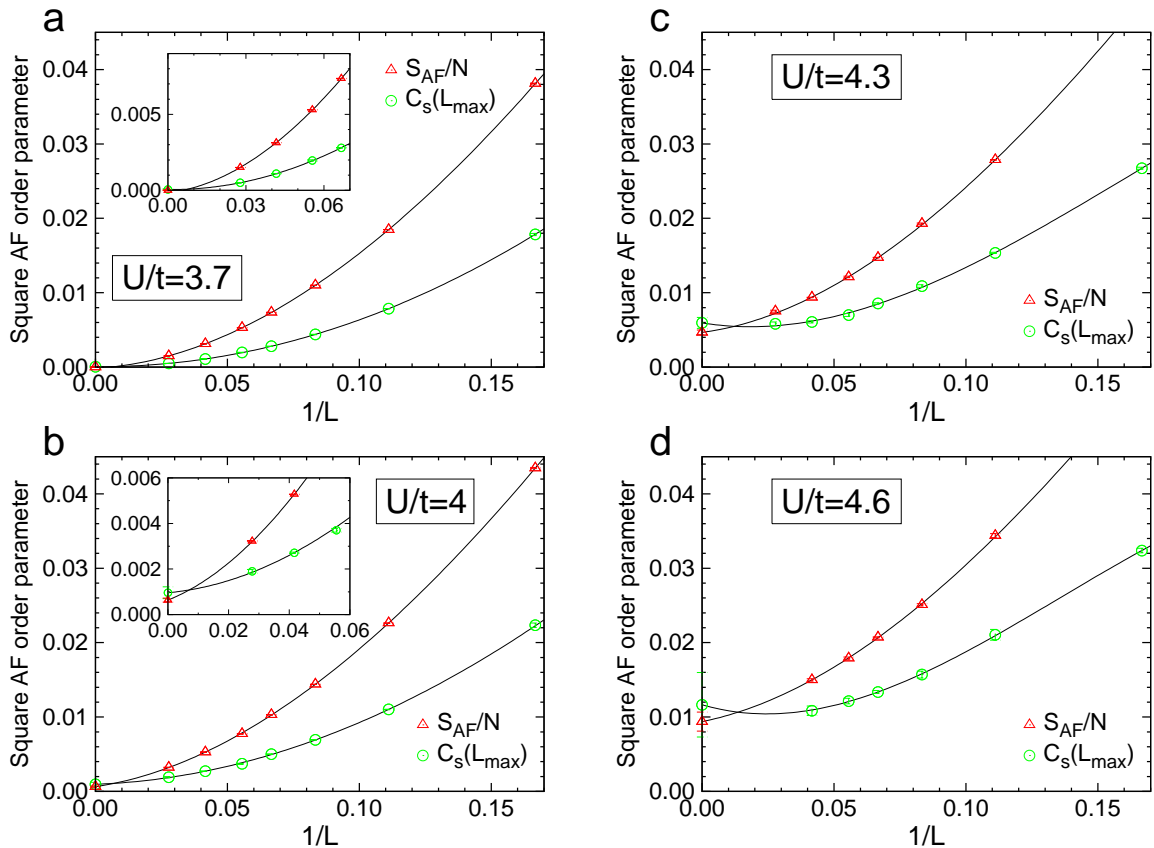}
\caption{\label{sq}
{\bf Finite size scaling of spin-spin correlation functions for the Hubbard model on the honeycomb lattice at half-filling.} 
Spin structure factor, $S_{\rm AF}$, and spin-spin correlations at the maximum distance, $C_s(L_{\rm max})$, are denoted 
by triangles and circles, respectively. Here, $L$ is the linear size of clusters containing $N=2L^2$ sites. 
Antiferromagnetic order parameter squared, $m_s^2$, is estimated by finite size extrapolating $S_{\rm AF}$ and $C_s(L_{\rm max})$ 
to $L\to\infty$, namely, $m_s^2=\lim_{L\to\infty}S_{\rm AF}/N=\lim_{L\to\infty}C_s(L_{\rm max})$. Solid curves are fit of the data by cubic polynomials in $1/L$.  It is clearly seen that a consistent extrapolated value $m_s^2$ is obtained for both quantities $S_{\rm AF}$ and 
$C_s(L_{\rm max})$, indicated respectively by triangles and circles at $1/L=0$. 
Error bars of the extrapolated values are computed with a resampling technique described in Methods.  
Insets show the expanded plots for large $L$. 
The fits are stable upon removal of the data for the largest (i.e., $L=36$) or the 
smallest (i.e., $L=6$) size, and the extrapolated value of $C_s(L_{\rm max})$ is always consistent with 
$S_{AF}/N$ within two standard deviations. 
Including the largest size calculations in the fits increases the extrapolated values slightly and at the same time gives 
more consistent values of $C_s (L_{\rm max})$ and $S_{AF}/N$ in $L\to\infty$, thus clearly indicating that our present 
estimate provides an accurate lower bound for the AF order parameter $m_s$. 
All data presented in this figure refers to $\Delta\tau t=0.1$, because the Trotter $\Delta \tau$  error is 
essentially negligible (see Fig.~\ref{phase_diagram}). 
More details are found in Supplementary information.  
}
\end{center}
\end{figure*}

First, we study both the spin structure factor $S_{\rm AF}={\frac{1}{N}} \langle\left[\sum_{\bm r}\left({\bm S}_{{\bm r},{\rm A}}-{\bm S}_{{\bm r},{\rm B}}\right)\right]^2\rangle$ and the spin-spin correlations 
$C_s({\bm R})=\langle {\bm S}_{{\bm r},{\rm A}}\cdot{\bm S}_{{\bm r}+{\bm R},{\rm A}}\rangle$ at the maximum distance 
$|{\bm R}|=L_{\rm max}$ of each cluster for $U/t=4$, where the strongest evidence of a spin liquid behavior was found 
in Ref.~\onlinecite{muramatsu}.  
Here ${\bm S}_{{\bm r},{\rm A}}$ (${\bm S}_{{\bm r},{\rm B}}$) is the spin operator at unit cell ${\bm r}$ on A (B) sublattice. 
As shown in Fig.~\ref{sq}{\bf b}, our results 
show consistently a finite value of the antiferromagnetic order parameter 
$m_s^2=S_{\rm AF}/N=C(L_{\rm max})$ for $L\to\infty$, in sharp contrast to the existence of a spin liquid, i.e., spin disordered, ground 
state reported in Ref.~\onlinecite{muramatsu}.

\begin{figure}
\begin{center}
\includegraphics[width=0.95\hsize]{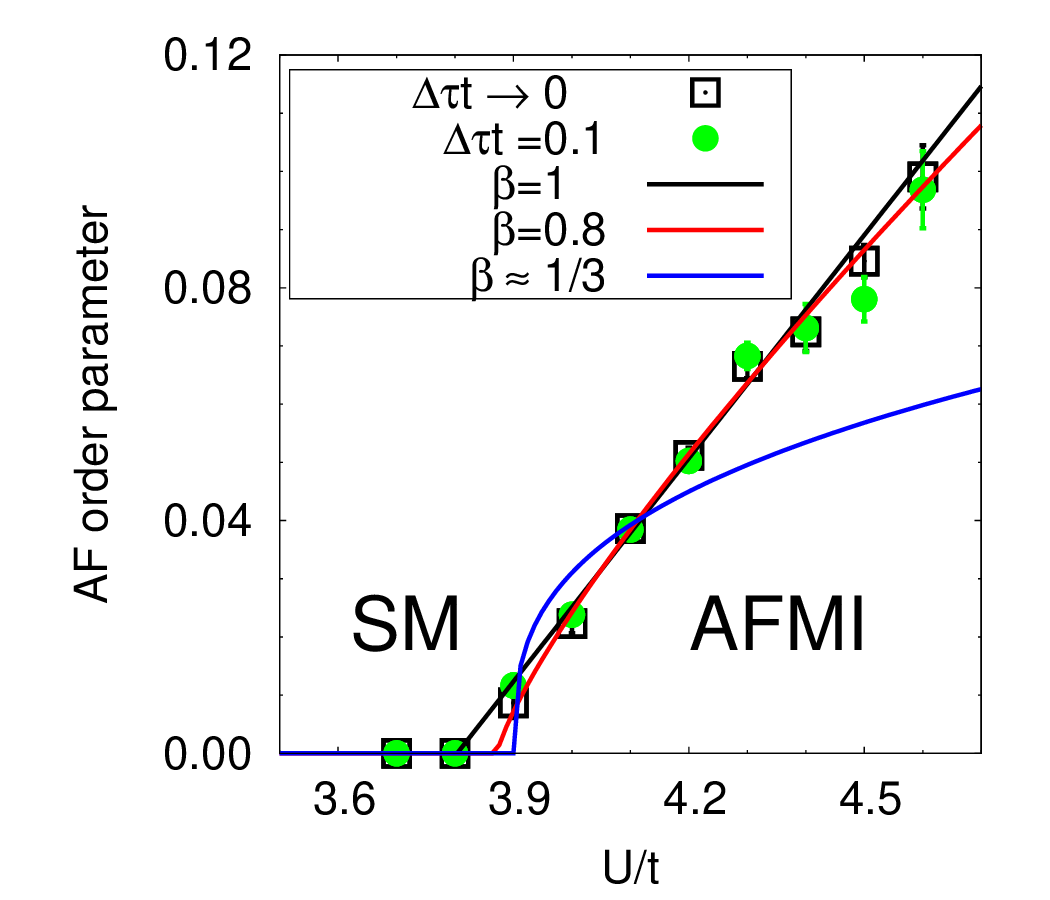} 
\caption{\label{phase_diagram}
{\bf The ground state phase diagram for the half-filled Hubbard model on the honeycomb lattice.} 
Antiferromagnetic order parameter $m_s$ (open squares) as a function of $U/t$. The  error, due to the 
finite $\Delta \tau$ in the evaluation of $S_{AF}$, is removed by quadratic extrapolations for $\Delta \tau t =0.1$, 
$\Delta \tau t =0.2$, and $\Delta \tau t =0.4$ (see Supplementary information for details). 
The antiferromagnetic order parameter $m_s$ is obtained by finite-size extrapolating 
the square root of $S_{AF}/N$, 
$m_s=\lim_{L\to\infty}\sqrt{S_{AF}/N}$, as shown in Fig.~\ref{sq}. 
For comparison, $m_s$ estimated by finite-size extrapolating $S_{AF}$ for $\Delta\tau t=0.1$ without the $\Delta\tau$ correction is 
also plotted (solid circles). 
SM and AFMI stand for semi-metal and antiferromagnetic insulator, respectively. 
Solid lines are fit of $m_s$ with the critical behavior $m_s = (U_c - U)^\beta$, for selected 
critical exponents $\beta$. $\beta=1$ for the HF theory~\cite{sorella_honey}, $\beta=0.3362$ for the 
classical critical theory of quantum magnets~\cite{brezin}, and $\beta=0.80\pm 0.04$ is 
the best fit of our data. 
In any case, the critical $U_c$ ranges from $U_c/t=3.8$ ($\beta=1$) to $U_c/t=3.9$ ($\beta=0.3362$). 
Our best estimate is $U_c/t=3.869\pm 0.013$.}
\end{center}
\end{figure}

By doing similar calculations for several $U/t$ values (see Fig.~\ref{sq} and Supplementary information), we find in 
Fig.~\ref{phase_diagram} that $m_s$ approximately scales linearly with respect to $U/t$, i.e., $m_s\propto |U-U_c|^\beta$, with a critical 
exponent $\beta \simeq 0.8$, which is similar to the critical behavior ($\beta=1$) predicted by the HF theory~\cite{sorella_honey}. 
Although corrections to this almost linear critical behavior are obviously expected, 
they do not change much the critical value $U_c$ at which the antiferromagnetic order melts, as clearly shown in Fig.~\ref{phase_diagram}. 
Our best estimate of the critical value is $U_c/t=3.869 \pm 0.013$, which is much smaller than the one ($\approx 4.3$) reported 
in Ref.~\onlinecite{muramatsu}. 
Note, however, that the critical exponent $\beta$ may be different from the present estimate if the critical region is very close 
to $U_c$. In such case  the accurate determination of $\beta$ obviously requires much larger clusters which are not feasible at present.

Let us now evaluate the spin gap $\Delta_s$. In order to avoid possible errors in extrapolating the imaginary time displaced spin-spin 
correlation functions, here we calculate directly the total energies in the singlet and the triplet sectors, with improved estimators, which 
dramatically reduce their statistical errors~\cite{hlubina} (also see Supplementary information). 
We can see clearly in Fig.~\ref{gap}{\bf a} that the extrapolated spin gaps for different $U/t$ values are always 
zero within statistical errors (e.g., the statistical error as small as $0.004t$ for $U/t=4$).

\begin{figure*}
\begin{center}
\includegraphics[width=0.92\hsize]{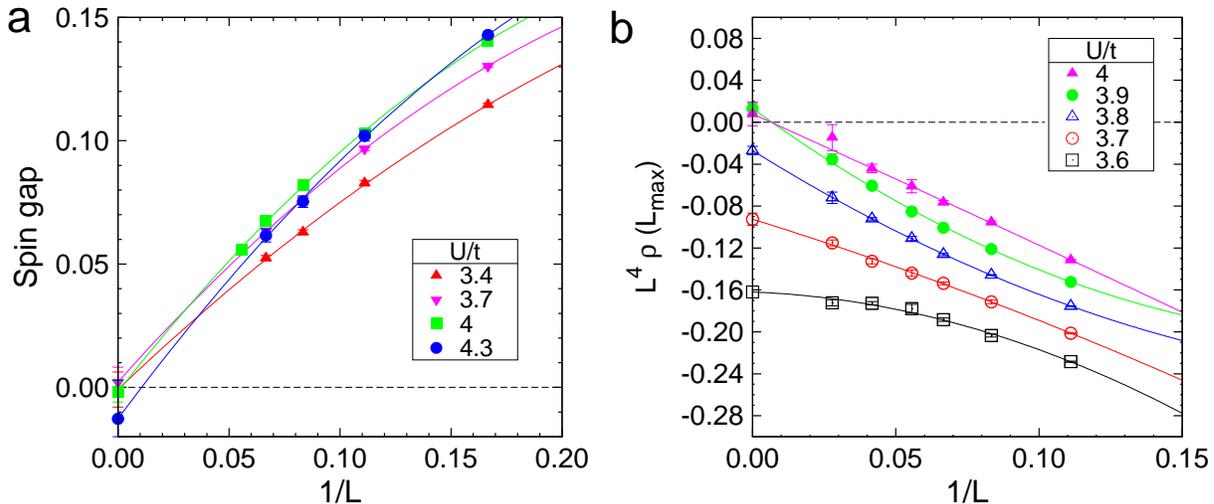}
\caption{\label{gap}
{\bf Finite size scaling of spin gap and charge-charge correlation functions for the Hubbard model on the honeycomb lattice at half-filling.}
{\bf a}, Spin gap $\Delta_s=E(S=1)-E(S=0)$ for various $U/t$, where 
$E(S)$ is the lowest energy for a given spin $S$. Solid curves are fits of data by quadratic 
polynomials in $1/L$. 
The extrapolated values are also indicated at $1/L=0$. Error bars of the extrapolated values are computed 
with the resampling technique. 
In the semi-metallic region, the spin gap scales to zero with increasing the 
resolution in momentum space, namely as $1/L$. In the antiferromagnetic region, the spin gap should instead vanish as $1/L^2$. 
This explains why for $U/t=4.3$ the gap extrapolates to negative values, as we are well inside the antiferromagnetic phase 
(see Fig.~\ref{phase_diagram}). 
In any case, a sizable spin gap is not found for any value of $U/t$. 
{\bf b}, Charge-charge correlation function 
$\rho({\bm R})=\langle n_{{\bm r},{\rm A}}n_{{\bm r}+{\bm R},{\rm A}}\rangle-\langle n_{{\bm r},{\rm A}}\rangle \langle n_{{\bm r}+{\bm R},{\rm A}}\rangle$ 
at the maximum distance $|{\bm R}|=L_{\rm max}$ 
for several values of $U/t$. In the semi-metallic phase, 
$\rho(R)\sim1/R^4$ and $L^4\rho(L_{\rm max})$ should converge to a finite value for $L\to\infty$. Instead, when a charge gap opens, 
the charge-charge correlations should decay exponentially and $L^4\rho(L_{\rm max})$ converges to zero in this limit. Indeed, a quadratic extrapolation to $L\to \infty$ of this quantity, which is clearly appropriate in the semi-metallic phase, 
appears to vanish in the interval between $U/t=3.8$ and $U/t=3.9$, in remarkable agreement with the critical value $U_c=3.869\pm 0.013$ 
estimated for the antiferromagnetic transition (see Fig.~\ref{phase_diagram}). 
Obviously, a polynomial fit is not consistent in the insulating region 
and this explains why the extrapolated value to $1/L=0$ seems slightly positive in this case.  
For the spin gap and the charge-charge correlation functions, the Trotter $\Delta \tau$ error is negligible, and all data shown here 
refers to $\Delta \tau t=0.14$ and $0.1$, respectively. 
}
\end{center}
\end{figure*}

Next, we investigate whether the system is metallic or insulating, namely, whether there exists a zero or a finite charge gap. For this purpose, 
it is enough to simply study the long distance behavior of charge-charge correlations, 
$\rho({\bm R})=\langle n_{{\bm r},{\rm A}}n_{{\bm r}+{\bm R},{\rm A}}\rangle-\langle n_{{\bm r},{\rm A}}\rangle \langle n_{{\bm r}+{\bm R},{\rm A}}\rangle$. 
Here $n_{{\bm r},{\rm A}}$ is the density operator at unit cell 
${\bm r}$ on A sublattice (see Fig~\ref{lattice}). They should change from power law to exponential behavior at a 
critical U where the charge gap opens up. 
This change of behavior is evident in Fig.~\ref{gap}{\bf b} and appears consistently around the onset of the 
antiferromagnetic transition ($U_c$), within a remarkably small uncertainty $<0.1t$ on the value of $U$. 
Our results therefore strongly support the 
more conventional scenario of a direct and continuous quantum phase transition between SM and AFMI~\cite{sorella_honey}.

\section*{Discussion}

Let us now discuss here why we have not found any evidence of a spin liquid phase. 
As shown in Ref.~\onlinecite{li}, by applying one of the theorems by Lieb~\cite{lieb2}, it is easily proved that 
the exact ground state of this model for $U\neq0$ satisfies the Marshall sign rule~\cite{sign} in the sector of no doubly occupied sites, 
accounting for low energy spin excitations. 
Indeed, the phases coincide with those of the simple antiferromagnetic N\'eel state ordered 
along the $x$-spin quantization axis, 
$\prod\limits_{{\bm R} \in A}  \left(|\uparrow\rangle_{\bm R} -|\downarrow \rangle_{\bm R}\right) \prod\limits_{{\bm R} \in B}  \left(|\uparrow\rangle_{\bm R} +|\downarrow \rangle_{\bm R}\right) $, 
where ${|\uparrow\rangle_{\bm R}}$ and ${|\downarrow\rangle_{\bm R}}$ are spin configurations (along the $z$-spin quantization axis) 
at site ${\bm R}$. The expansion of this state in terms of ${|\uparrow\rangle_{\bm R}}$ and ${|\downarrow\rangle_{\bm R}}$ yields 
the simple Marshall sign, namely, it is negative if the number of spin down in the A sublattice is odd. 
Thus, the phases of the ground state are trivial in the bosonic spin 1/2 sector. 
Therefore, Bose-Einstein condensation can hardly  be avoided and a magnetic long-range order occurs once the charge gap becomes finite. 

At this point, one could be tempted to assume the general validity of the above observation 
for generic S=1/2 model Hamiltonians with SU(2) invariance and use this criteria based on the phases of the 
ground state wave function  as a powerful guideline in the search of spin liquids for model systems as well as for real materials. 
Indeed, in all unfrustrated spin-$1/2$ Heisenberg and 
Hubbard models in the sector of no doubly occupied sites, the phases of the ground state wave function are not at all entangled in 
real space as they factorize into independent contributions relative to each site. 
Instead, the phases of the ground state wave functions are highly non trivial in  well established spin liquid models 
such as, for instance, the Kitaev's model~\cite{kitaev}, and the celebrated quantum dimer model on the triangular lattice~\cite{triang}, 
because they are described by paired wave functions, which couple in a non trivial way 
the phases of nearest neighbor spins~\cite{yunoki}. 

Therefore, we conclude  that in a true spin liquid in 2D,  the phases of the ground state wave function should be highly non trivial 
and entangled, otherwise any seed of spin liquid behavior would be most likely destabilized.
To our best knowledge, the above observation is valid so far for all spin-$1/2$ models 
with $SU(2)$ invariance. Notice that the restriction to  SU(2) invariant models appears 
to be important because the spin-1/2 easy-axis Heisenberg model on the 
Kagome lattice most likely display spin liquid behavior \cite{balents3,melko,melko2}. 
Here, however, the calculations have not been confirmed on fairly low temperatures yet. 
Therefore, further numerical study is required for understanding what are the key
 ingredients  that stabilize a spin liquid phase in "realistic" electronic models.

\section*{Methods}
Here we employ the standard auxiliary field Monte Carlo algorithm~\cite{hirsch} with a more efficient implementation~\cite{hlubina} 
by using different left and right trial functions $|\psi_L\rangle$ and $|\psi_R\rangle$ in equation~(\ref{ave}). 
We include also in the trial wave function a Gutzwiller type projection, 
$\exp (-g \sum_i n_{i\uparrow}n_{i\downarrow})$, where $g$ is the Gutzwiller variational parameter, to optimize the efficiency. 
As reported in Ref.~\onlinecite{hlubina}, the statistical error in evaluating the 
energy $E(S)$ for a given spin $S$ is dramatically reduced for appropriate values of $g$. Thus we can evaluate the 
spin gap $\Delta_S= E(S=1) - E(S=0)$ with high accuracy, without facing the negative sign problem, by directly 
simulating the singlet $S=0$ and the triplet $S=1$ sectors separately (see Supplementary information). 

In order to accelerate the convergence to the ground state, we use for 
$|\psi_R\rangle$ a Slater determinant with a definite spin $S$, 
by breaking only spatial 
symmetries to remove the degeneracy at momenta $K$ and $K'$ 
for the clusters chosen 
(a similar strategy was adopted in Ref.~\onlinecite{muramatsu}). 
Conversely, we use for $|\psi_L\rangle$ a rotational and translationally invariant Slater determinant obtained by diagonalizing 
a mean field Hamiltonian, containing an explicit antiferromagnetic order parameter directed along the $x$-spin quantization 
axis. In this way, the left and the right trial wave functions break different symmetries (spin and spatial ones, respectively), 
and for any symmetric 
operator ${\hat O}$ the convergence to the ground state is expected to be much faster because it is dominated by the singlet  gap in the symmetric sector 
$\Delta_{\rm gap}$, that is clearly much larger than, e.g., the  lowest triplet 
excitation in the magnetic phase. 
Since $\Delta_{\rm gap}$ is expected to scale to zero (if indeed zero) at most as $\simeq 1/L$, we use a projection time 
$\tau = (L+4) /t$, which we have tested carefully to be large enough for well converged results (see Supplementary information). 
We have also checked that the systematic error due to discretizing $\tau$ is basically negligible with $\Delta\tau t =0.14$ for the spin gap 
calculations and with $\Delta\tau t =0.1$ for the correlation functions (see Fig.~\ref{phase_diagram} and Supplementary information).

In order to evaluate the statistical errors of the finite size extrapolations, we use a 
straightforward resampling technique. 
This resampling technique is used, for example, when values of $S_{AF}$ calculated for finite sizes are extrapolated to $L\to\infty$ 
to estimate $m_s$ in Fig.~\ref{phase_diagram}. Let us denote in general the calculated Monte Carlo data $f(L)$ and the corresponding 
statistical error $\delta f(L)$ obtained for a cluster of size $L$. The main goal of this 
resampling technique is to estimate the finite-size extrapolated value $c_0$ and its statistical error $\delta c_0$ when the Monte Carlo 
data are fitted by, e.g., cubic polynomials, 
\begin{equation}
f(L)=\sum_{n=0}^3{\frac{c_n}{L^n}}.
\label{fit}
\end{equation} 
In this resampling technique, we first generate for each $L$ a "fictitious sample" which is normally distributed around $f(L)$ with 
its standard deviation $\delta f(L)$, which is also an output of the quantum Monte Carlo simulation. 
Then, we fit these "fictitious" data to equation (\ref{fit}), by using the weighted 
(with $1/\left(\delta f(L)\right)^2$) least square fit, and estimate $c_0$. We repeat this $M_{\rm rs}$ 
times so that we have now $M_{\rm rs}$ numbers of samples for $c_0$, i.e., $\{c_0^{(1)}, c_0^{(2)}, \dots, c_0^{(M_{\rm rs})}\}$ 
distributed according to a probability distribution consistent with the Monte Carlo simulations. 
Finally, we simply average $\{c_0^{(i)}\}$ ($i=1,2,\dots,M_{\rm rs}$) for $\langle c_0\rangle ={\displaystyle \lim_{L\to\infty}}f(L)$, 
and the standard deviation of $\{c_0^{(i)}\}$ gives an estimate of the statistical error $\delta c_0$ of the extrapolated value 
$\langle c_0\rangle$. We take $M_{\rm rs}=200$. We have checked that the resultant $\langle c_0\rangle$ and $\delta c_0$ are 
not dependent on $M_{\rm rs}$ as long as $M_{\rm rs}$ is large enough.

\begin{acknowledgments}
We acknowledge E. Tosatti, F. Becca, and T. Li~for useful discussions. We are also grateful to A.~Muramatsu 
and F.~F.~Assaad for valuable comments and providing us some of their numerical data reported in Ref.~\onlinecite{muramatsu}. 
This work is supported by a PRACE grant 2010PA0447 and by MIUR-COFIN2012. 
Part of the results is obtained by 
the K computer at RIKEN Advanced Institute for Computational Science. 
\end{acknowledgments}

\section*{Additional information}
The authors declare no competing financial interests. Supplementary information accompanies this paper on www.nature.com/naturephysics. 
Reprints and permissions information is available online at www.nature.com/scientificreports. 
Correspondence and requests for materials should be addressed to S.S or S.Y.

\newpage

\title{Absence of a Spin Liquid Phase in the Hubbard Model on the Honeycomb Lattice}

\author{Sandro Sorella$^{1,2,3}$}
 \email{sorella@sissa.it}
\author{Yuichi Otsuka$^{3}$}
 \email{otsukay@riken.jp}
\author{Seiji Yunoki$^{3,4,5}$}
 \email{yunoki@riken.jp}
\affiliation{%
$^1$SISSA -- International School for Advanced Studies, Via Bonomea 265, 34136 Trieste, Italy, \\
$^2$Democritos Simulation Center, CNR -- IOM Instituto Officina dei Materiali, 34151 Trieste, Italy, \\
$^3$Computational Materials Science Research Team, RIKEN AICS, Kobe, Hyogo 650-0047, Japan, \\
$^4$Computational Condensed Matter Physics Laboratory, RIKEN ASI, Saitama 351-0198, Japan, and \\
$^5$CREST, Japan Science and Technology (JST), Kawaguchi, Saitama 332-0012, Japan
}%

\maketitle

\section*{Supplementary information}

In this Supplementary information, we explain details of the trial wave functions used and provide numerical data for correlation 
functions and spin gap of the half-filled Hubbard model on the honeycomb lattice. The Hubbard model is described by the following 
Hamiltonian: 
\begin{equation}
{\hat H} = -t \sum_{\langle{\bm R},{\bm R'}\rangle,\sigma}\left(c^\dag_{{\bm R},\sigma}c_{{\bm R'},\sigma} +{\rm h.c.}\right)
+U\sum_{\bm R}n_{{\bm R},\uparrow}n_{{\bm R},\downarrow}, 
\end{equation} 
where $c^\dag_{{\bm R},\sigma}$ is an electron creation operator at site ${\bm R}$ with spin $\sigma=(\uparrow,\downarrow)$, 
$n_{{\bm R},\sigma}=c^\dag_{{\bm R},\sigma}c_{{\bm R},\sigma}$, and $\langle{\bm R},{\bm R'}\rangle$ runs over all nearest neighbor 
sites ${\bm R}$ and ${\bm R'}$. Here the electron hopping is described by the first term (${\hat H}_0$) and the one-site Coulomb 
interaction is represented by the second term (${\hat H}_{\rm I}$).  

\section{Honeycomb lattice}
 
\begin{figure}[htb]
\centering
\includegraphics[width=0.9\hsize]{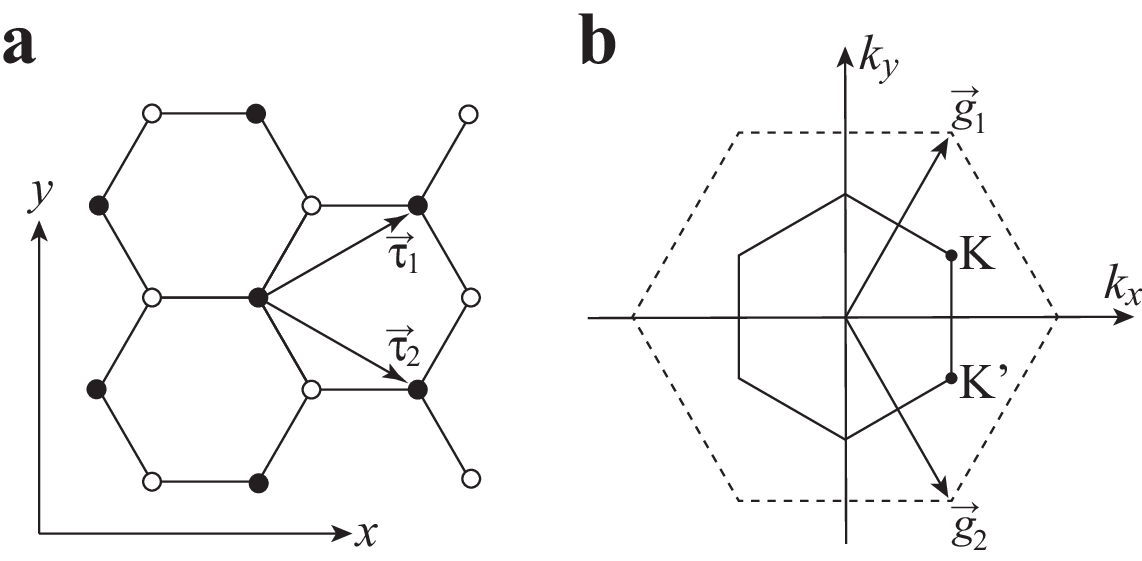}
\caption{
\label{bz}
(a) The honeycomb lattice with the primitive lattice vectors $\vec\tau_1=(3/2, \sqrt{3}/2 )$ and $\vec\tau_2=(3/2, -\sqrt{3}/2 )$. 
Each unit cell contains two sites (indicated by open and solid circles) belonging to A and B sublattices. (b) The first Brillouin zone 
(denoted by solid lines forming hexagon) in the momentum space with the reciprocal lattice vectors $\vec g_1=2\pi(1/3, 1/\sqrt{3} )$ and 
$\vec g_2=2\pi(1/3, -1/\sqrt{3} )$. Two high symmetric and independent momenta are indicated by K: $2\pi(1/3, 1/3\sqrt{3} )$ 
and K': $2\pi(1/3, -1/3\sqrt{3} )$.  K and K' are allowed only when $L$ is a multiple of 3 for a finite lattice of $N=2L^2$ with periodic 
boundary conditions. 
}
\end{figure}

As shown in Fig.~\ref{bz}{\bf a}, the honeycomb lattice is formed by the primitive lattice vectors 
(with lattice constant $a=1$):
\begin{eqnarray}
\vec \tau_1 &=& (3/2, \sqrt{3}/2 )\\
\vec \tau_2 &=& (3/2, - \sqrt{3}/2 ). 
\end{eqnarray}
Each unit cell contains two sites, belonging to different sublattices (A and B sublattices), and using $\vec\tau_1$ and $\vec\tau_2$ 
these sites are defined by 
\begin{eqnarray}
{\bm R}_A&=& n \vec \tau_1 + m \vec \tau_2  \\
{\bm R}_B&=& (n -1/3) \vec \tau_1  + (m -1/3) \vec \tau_2,  \\
\end{eqnarray}
where $n$ and $m$ are integers.
Periodic boundary conditions are obtained simply by requiring equivalence of 
lattice points when they differ by the lattice vectors ${\bm T}_1$ and ${\bm T}_2$ which define the lattice: 
\begin{eqnarray}
\bm T_1 &=& L \vec \tau_{1} \\
\bm T_2 &=& L \vec \tau_{2}. 
\end{eqnarray}
The lattice thus contains $N=2L^2$ independent sites, and satisfies all point group symmetries of the infinite lattice~[\onlinecite{bernu_sup}]. 
We have chosen $L$ to be a multiple of $3$. For this reason, two sites at the maximum distance $L_{\max}=L$ in a given finite lattice are 
in the same sublattice, and are separated by a lattice vector 
\begin{equation}
\vec \tau_{\rm max}={\frac{L}{3}} (\vec\tau_1  + \vec\tau_2).  
\end{equation}
This maximum distance vector differs slightly from the one defined in Ref.~\onlinecite{muramatsu_sup} because 
here we consider also the rotation symmetry of 120 degrees around 
each site. This symmetry is important in order to define properly the maximum distance possible in a given finite lattice 
among all equivalent ones. 

The reciprocal lattice vectors in momentum space are given by 
\begin{eqnarray}
\vec g_1&=&2\pi(1/3, 1/\sqrt{3} )\\
\vec g_2&=&2\pi(1/3, -1/\sqrt{3} ), 
\end{eqnarray}
and using these vectors a momentum $\vec k$ in a given finite lattice of $N=2L^2$ is described by 
\begin{equation}
\vec k = {\frac{k_1}{L}}\vec g_1 + {\frac{k_2}{L}}\vec g_2,  
\end{equation}
where $k_1,k_2=0,1,\dots,L-1$. Thus, only when $L$ is a multiple of $3$, the high symmetric momenta K: $2\pi(1/3, 1/3\sqrt{3} )$ and 
K': $2\pi(1/3, -1/3\sqrt{3} )$ are allowed (see in Fig.~\ref{bz}{\bf b}). These are the momenta where Dirac points appear and in our study 
we choose $L$ to be a multiple of 3 for the systematic finite-size scaling analysis.  

\section{Trial wave functions}

\begin{figure*}[htb]
\centering
\includegraphics[width=0.8\hsize]{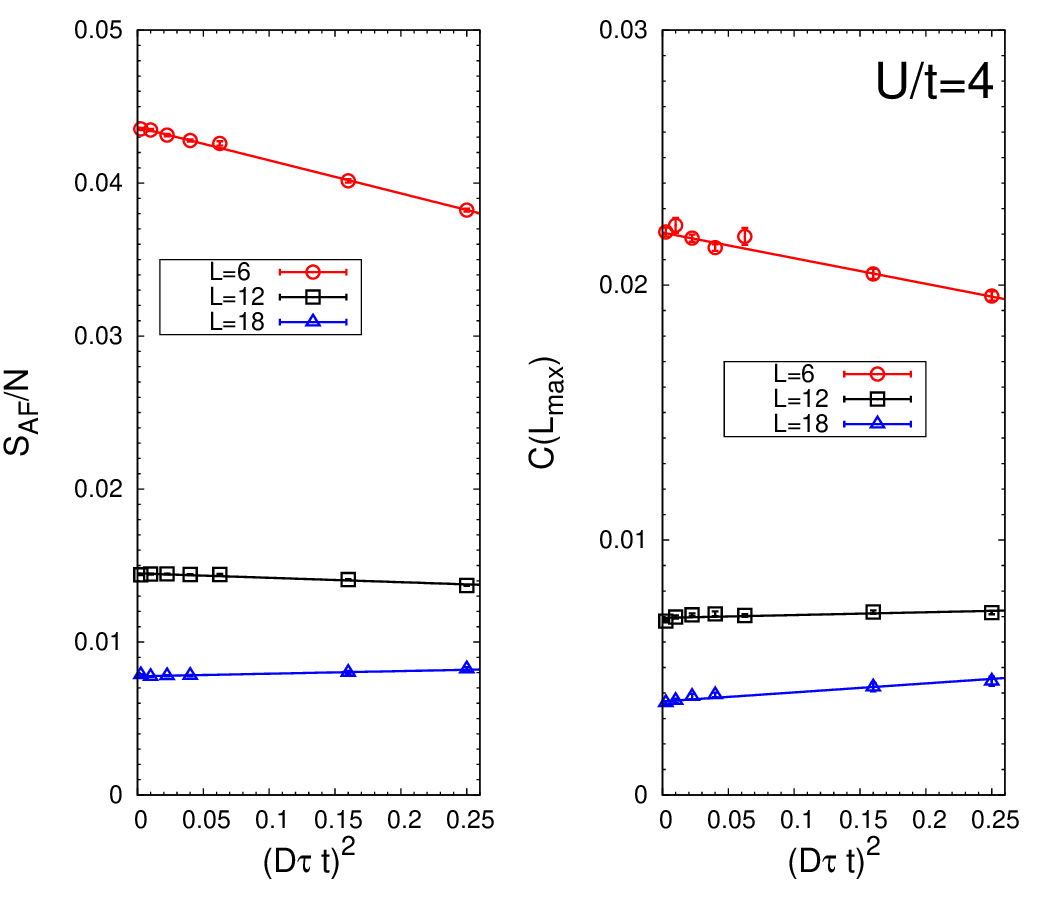}
\caption{
\label{plotall}
Spin structure factor, $S_{\rm AF}$, and spin-spin correlation functions at the maximum distance, 
$C(L_{\rm max})$, as a function of the time discretization $\Delta \tau $ used in the 
Trotter decomposition in Eq.~(\ref{decomp}) for $U/t=4$. The system size $L$ is indicated in the figures. The lines 
are linear fits of data in $(\Delta\tau)^2$. 
}
\end{figure*}

We use the auxiliary field quantum Monte Carlo technique~\cite{hirsch_sup} to statistically evaluate 
\begin{equation}
O(\tau)={ \langle \psi_L |  {\rm e}^{-{\frac{\tau}{2}} {\hat H} } {\hat O} {\rm e}^{ - {\frac{\tau}{2}} {\hat H} } | \psi_R \rangle 
\over  \langle \psi_L | {\rm e}^{ - \tau {\hat H} } | \psi_R \rangle },  
\label{ave}
\end{equation}
where ${\hat O}$ is a physical operator and $\tau$ is the projection time. $|\psi_R \rangle $ and $|\psi_L\rangle $ are respectively the 
right and the left trial wave functions, which are not orthogonal to the exact ground state wave function.  
The exact ground state expectation value $\langle{\hat O}\rangle$ of the operator ${\hat O}$ is then 
obtained by adopting the limit of $\tau\to \infty$ and $\Delta\tau \to 0$ for $O(\tau)$, where $\Delta\tau$ is the short time 
discretization of $\tau$ introduced in the Suzuki-Torotter decomposition~\cite{suzuki_sup}  
\begin{equation}
{\rm e}^{-\tau\hat H}=\left[ {\rm e}^{-{\frac{1}{2}}\Delta\tau\hat H_0} {\rm e}^{-\Delta\tau\hat H_{\rm I}}  {\rm e}^{-{\frac{1}{2}}\Delta\tau\hat H_0} +{\cal O}\left( \left(\Delta\tau\right)^3\right) \right]^{N_\tau} 
\label{decomp}
\end{equation}
with $\tau=N_\tau \Delta\tau$. 
As shown in Fig.~\ref{plotall}, the systematic error introduced in this decomposition is negligible for small 
$\Delta\tau$. 
Notice also that the decomposition adopted here is symmetric, thus allowing an $O(\Delta \tau^3)$ error in the short time propagation 
(the non-symmetric form, ${\rm e}^{-\Delta\tau {\hat H}}\approx{\rm e}^{-\Delta\tau {\hat H}_0}{\rm e}^{-\Delta\tau {\hat H}_{\rm I}}$, has 
a much larger error $O(\Delta \tau^2)$).
Therefore, we can use larger $\Delta\tau$ with 
excellent accuracy. Indeed, we find that $\Delta\tau t=0.14$ ($0.1$) is small enough to ignore the systematic errors for 
the spin gap (spin and charge correlation) calculations.   

In our calculations, the left trial  function $|\psi_L\rangle$ is given by 
\begin{equation}
|\psi_L\rangle={\hat P}_{\rm G}|\Phi_{\rm MF}\rangle,
\label{wf_left}
\end{equation}
where ${\hat P}_{\rm G}$ is the Gutzwiller projection 
\begin{equation}
{\hat P}_{\rm G} = \prod_{\bm R}\exp(-gn_{{\bm R},\uparrow}n_{{\bm R},\downarrow}),
\end{equation}
and $|\Phi_{\rm MF}\rangle$ is the $N$-electron ground state of the following mean field Hamiltonian:
\begin{eqnarray} \label{mf}
{\hat H}_{\rm MF} &=& -\sum_{\langle {\bm R},{\bm R}'\rangle}\sum_\sigma  \left( c^{\dag}_{{\bm R}, \sigma} c_{{\bm R}', \sigma} + {\rm h.c.} \right) \nonumber \\
&+&
\Delta \sum_{\bm R} (-1)^{s({\bm R})} \left( c^{\dag}_{{\bm R},\uparrow} c_{{\bm R},\downarrow} + c^{\dag}_{{\bm R},\downarrow} c_{{\bm R},\uparrow}  \right), 
\end{eqnarray}
with $\Delta$ being the antiferromagnetic order parameter, and $s({\bm R})=0$ (1) if site ${\bm R}$ belongs to the A (B) sublattice. 
Obviously, $|\psi_L\rangle$ breaks the $SU(2)$ spin rotational symmetry, but conserves the spatial 
rotational symmetry. In order to obtain a rapid  convergence with respect to $\tau$ in Eq.~(\ref{ave}), we use optimized values 
of $g$ and $\Delta$ which maximize $\langle \psi_L | {\rm e}^{ - \tau {\hat H} } | \psi_L \rangle / \langle \psi_L | \psi_L \rangle$.

For the right trial wave function $|\psi_R\rangle$, we simply use a Slater determinant constructed by the single-particle states of 
the non-interacting $\hat H_0$ (with eigenstates $\varepsilon_{\bm k}$), occupied from the lowest energy states by 
$N_\uparrow$ number of up electrons and $N_\downarrow$ number of down electrons with 
$N_\uparrow = N_\downarrow = N/2$ for the spin singlet ($S=0$) and 
$N_\uparrow-1 = N_\downarrow+1 = N/2$ for the spin triplet ($S=1$). 
In order to remove the open shell condition in the single-particle energy level $\varepsilon_{\bm k}$ at the Fermi level and thus to 
avoid the negative sign problem, we add to $\hat H_0$ an appropriate 
small perturbative term. 
For example, to remove the degeneracy at momenta $K$ and $K'$, a tiny perturbation term $\delta t\sum_{{\bm R}_A,\sigma}\left( c^\dag_{{\bm R}_A,\sigma}c_{{\bm R}_A-{\vec\tau_{AB}},\sigma}+{\rm h.c.}\right)$ with $\vec\tau_{AB}={\frac{1}{3}}(\vec\tau_1+\vec\tau_2)$ is added 
to $\hat H_0$. This term certainly breaks the spatial rotational symmetry, but preserve exactly the spin rotational symmetry.

The left and the right trial wave functions therefore break different symmetries (spin and spatial rotations, respectively). An 
advantage of using these trial wave functions is that for any symmetric operator ${\hat O}$ the convergence to the ground state is expected to be much faster in $\tau$ 
because it is dominated by the singlet gap in the symmetric sector 
$\Delta_{\rm gap}$, that is clearly much larger than, e.g., the  lowest triplet 
excitation in the magnetic phase.
Since $\Delta_{\rm gap} $ is expected to scale to zero (if indeed zero) at most as $\simeq 1/L$, we take a projection time 
$\tau = (L+4) /t$, which, as shown below, is large enough for well converged results.

It should be emphasized that with these trial wave functions we are able to perform the quantum Monte Carlo simulations for the spin singlet 
and the spin triplet sectors independently with no negative sign problem. 
This is one of the crucial technical points in our simulations since we can estimate 
the spin gap directly and accurately. This should be contrasted to Ref.~\onlinecite{muramatsu_sup} where the estimation of the spin 
gap relies on the asymptotic behavior of the imaginary time displaced spin-spin correlation functions, which is sometime very difficult to 
extract accurately.

In addition, we use a ''mixed average'' for the energy calculations, i.e., 
\begin{equation}
\langle \hat H \rangle = \lim_{\tau\to\infty} { \langle \psi_L |  \hat H {\rm e}^{-\tau\hat H} | \psi_R \rangle 
\over  \langle \psi_L | {\rm e}^{ - \tau {\hat H} } | \psi_R \rangle }.  
\end{equation}
With this mixed average, we can significantly reduce the statistical error as compared to the one obtained with Eq.~(\ref{ave}). The reduction 
of the statistical error is simply because this mixed average satisfies the zero variance principle,
namely that the statistical error is zero 
if $|\psi_L\rangle$ is exact.

\section{Correlation functions and spin gap}

The spin-spin correlation functions at the maximum distance $C_s(L_{\rm max})$ is defined by 
\begin{equation}
C_s(L_{\rm max})={\frac{1}{NN_{\vec\tau_{\rm max}}}}\sum_{{\bm R},\vec\tau_{\rm max}}\langle {\bm S}_{\bm R}\cdot {\bm S}_{{\bm R}+\vec\tau_{\rm max}}\rangle, 
\end{equation}
where ${\bm S}_{\bm R}$ is the spin operator at site ${\bm R}$, $\vec\tau_{\rm max}$ runs over all symmetrically equivalent 
maximum distance vectors, $N_{\vec\tau_{\rm max}}$ is the number of these vectors, and 
$L_{\rm max}=|\vec\tau_{\rm max}|$~\cite{notemuramatsu_sup}. As mentioned above, spins at site ${\bm R}$ and site 
${\bm R}+\vec\tau_{\rm max}$ are in the same sublattice. We also study the spin structure factor, which is defined by 
\begin{equation}
S_{\rm AF} = {\frac{1}{N}} \left\langle  \left[ \sum_{\bm r} \left( {\bm S}_{{\bm r},A} - {\bm S}_{{\bm r},B} \right) \right]^2 \right\rangle,
\end{equation}
where ${\bm S}_{{\bm r},A}$ and ${\bm S}_{{\bm r},B}$ are the spin operators at unit cell ${\bm r}$ on A and B sublattices, 
respectively~\cite{notemuramatsu_sup}. 

\begin{figure*}[htb]
\centering
\includegraphics[width=0.9\hsize]{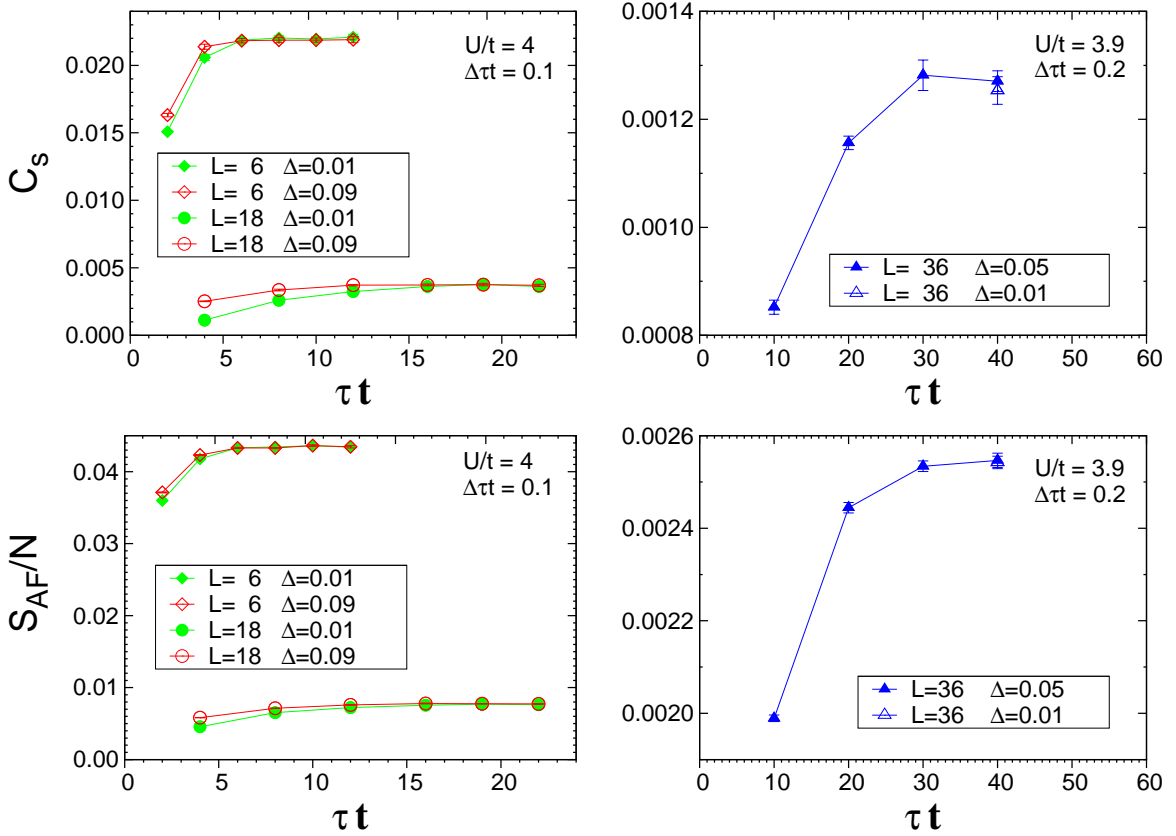}
\caption{
\label{spectrum}
The spin-spin correlations at the maximum distance $C_s(L_{\rm max})$ (upper panels) and 
magnetic structure factor $S_{AF}/N$ (lower panels) as a function of the  projection time $\tau$ 
for $L=6$, $18$, and $L=36$. $U/t$, $\Delta\tau t$, and $\Delta$ used are indicated in the figures.  
The converged results are obtained at $\tau t= L+4$, regardless of the chosen value of the parameter $\Delta$ defining the left trial function 
$|\psi_L\rangle$ in Eq.~(\ref{mf}). Notice that, for $\Delta$ larger  than  $0.01$, 
the convergence to the ground state vales ($\tau \to \infty$) is apparently faster. 
The Gutzwiller parameter $g$ in $|\psi_L\rangle$ is set to be $0.72$ for $U/t=4$ and $L=6, 18$, and 
$0.65$ for $U/t=3.9$ and $L=36$. 
Notice also that these antiferromagnetic correlation functions,  $C_s(L_{\rm max})$ and $S_{AF}/N$, always monotonically 
increase within statistical errors with the projection time $\tau$, 
and converge even in the most difficult cases (right panels), namely, for the largest size very close to the critical 
$U$ ($U_c/t\sim 3.87$) below which the antiferromagnetic order disappears. 
}
\end{figure*}

We first show in Fig.~\ref{spectrum} the projection time $\tau$ dependence of $C_s(L_{\rm max})$ and $S_{\rm AF}$ for $L=6$ and 
$18$ with $U/t=4$, and for $L=36$ with $U/t=3.9$, which is very close to $U_c/t\sim3.87$. Here, two different values of 
the antiferromagnetic order parameter $\Delta$ in ${\hat H}_{\rm MF}$ [Eq.~(\ref{mf})] are chosen for the left trial function 
$|\psi_L\rangle$ described by Eq.~(\ref{wf_left}).  
We can clearly see in Fig.~\ref{spectrum} that (i) both $C_s(L_{\rm max})$ and $S_{\rm AF}$ are well converged at $\tau t=L+4$ 
regardless of $\Delta$ values and (ii) for the chosen $\Delta$'s the convergence of both quantities is always monotonically increasing 
within statistical errors, clearly showing that a finite projection time $\tau$ can at most underestimate the magnetic order parameter.

TABLES~\ref{clmaxdt.4}--\ref{clmaxextr} (\ref{spidt.4}--\ref{spiextr}) summarize 
$C_s(L_{\rm max})$ ($S_{\rm AF}/N$) calculated for different system sizes for several 
values of $U/t$. 
The thermodynamically extrapolated values for $L\to\infty$ are also shown in these tables. 
The extrapolation is performed by fitting the calculated data with cubic polynomials in $1/L$, and the 
simple resampling method, explained in the Methods of the main text, is employed to 
estimate the statistical error of the extrapolated value. 
To check the stability of this finite size extrapolation, we carry out three different cubic polynomial fitting, 
whenever possible, namely, by using (i) all data for $L=6$, $12$, $18$, $24$, and $36$, 
(ii) all data but without the largest size data ($L=36$), and (iii) all data 
but without the smallest size data ($L=6$). The results are shown in the last three rows of TABLES~\ref{clmaxdt.4}--\ref{spiextr}. 
We can see in these table that these three different fits give quantitatively the same extrapolated results within the statistical 
error, strongly indicating that our fitting is indeed stable. 

The antiferromagnetic order parameter $m_s$ is obtained by $m_s=\sqrt{S_{\rm AF}/N}=\sqrt{C(L_{\rm max})}$ for $L\to\infty$. 
Although $C_s(L_{\rm max})$ and $S_{\rm AF}$ are certainly different for a given finite $L$, the thermodynamically 
extrapolated values, namely $m_s^2$, are statistically consistent, as clearly seen in TABLES~\ref{clmaxdt.4}--\ref{spiextr}. 
We also find that the general trend of the extrapolation
in the thermodynamic limit $L \to \infty$ is that 
the order parameter $m_s$ estimated slightly increases if we remove the smallest sizes in the fit.  
This trend is also particularly evident even when quadratic polynomial is used for fitting data (not reported).
This suggests that our estimated values $m_s$ plotted in Fig. 3 of the main text 
and the ones reported in TABLES~\ref{clmaxdt.4}--\ref{spiextr} 
can be considered to be accurate lower bounds of the order parameter.

\begin{table*}[htb]
\begin{tabular}{|l|c|c|c|c|c|} \hline
\backslashbox{$L$}{$U/t$} & 3.7 & 3.8 & 3.9 & 4   & 4.1\\ \hline
6               & 0.01769(73)       & 0.018160(40)      & 0.019278(70)      & 0.02044(20)       & 0.021748(82)     \\
9               & 0.007940(60)      & 0.008924(55)      & 0.00999(11)       & 0.01102(13)       & 0.012099(67)     \\
12              & 0.004652(35)      & 0.005429(15)      & 0.006229(46)      & 0.007263(32)      & 0.008307(30)     \\
15              & 0.003033(19)      & 0.003787(26)      & 0.00491(23)       & 0.005485(41)      & 0.006541(47)     \\
18              & 0.002235(28)      & 0.002840(17)      & 0.003360(85)      & 0.004419(27)      & 0.005536(47)     \\
24              & 0.001383(28)      & 0.001921(12)      & 0.002437(60)      & 0.003325(37)      & 0.004466(89)     \\
36              & 0.000674(21)      & 0.0011659(73)     & 0.001658(82)      & 0.002380(67)      & 0.003487(49)     \\
$\infty$ (i)    & -0.00011(15)      & 0.000384(55)      & 0.00120(35)       & 0.00125(25)       & 0.00241(20)      \\
$\infty$ (ii)   & 0.00048(39)       & 0.00076(19)       & 0.00155(65)       & 0.00148(43)       & 0.00332(56)      \\
$\infty$ (iii)  & -0.00018(20)      & 0.000264(95)      & 0.00098(56)       & 0.00102(38)       & 0.00184(35)      \\
\hline \hline
\backslashbox{$L$}{$U/t$} & 4.2 & 4.3 & 4.4 & 4.5 & 4.6\\ \hline
6               & 0.023026(97)      & 0.02468(11)       & 0.025896(80)      & 0.027285(89)      & 0.02818(68)      \\
9               & 0.013398(78)      & 0.014594(91)      & 0.01609(16)       & 0.01771(17)       & 0.01864(19)      \\
12              & 0.009512(41)      & 0.01071(20)       & 0.012258(85)      & 0.01382(30)       & 0.01424(27)      \\
15              & 0.007674(55)      & 0.00894(12)       & 0.01042(34)       & 0.01150(22)       & 0.0124(03)       \\
18              & 0.006687(85)      & 0.00739(27)       & 0.00893(21)       & 0.01145(91)       & 0.0115(04)       \\
24              & 0.00565(16)       & 0.00731(58)       & 0.00864(61)       & 0.0109(11)        & 0.0098(12)       \\
36              & 0.00457(19)       & 0.00663(53)       &  ---              & 0.00731(78)       &  ---             \\
$\infty$ (i)    & 0.00406(53)       & 0.0067(15)        &  ---              & 0.0061(22)        &  ---             \\
$\infty$ (ii)   & 0.00510(84)       & 0.0051(28)        & 0.0057(23)        & 0.0141(53)        & 0.0120(49)       \\
$\infty$ (iii)  & 0.00308(87)       & 0.0084(25)        &  ---              & 0.0048(40)        &  ---             \\
\hline
\end{tabular}
\caption{%
Spin-spin correlations at the maximum distance $C_s(L_{\rm max})$ 
for several values of $U/t$.
Here $\Delta\tau t=0.4$ and projection time $\tau t=L+4$. 
The extrapolated values for $L\to\infty$ are obtained 
by cubic fits in $1/L$ using 
(i)  data corresponding to $L=6,9,12,15,18,24,36$, 
(ii) data as in (i) but without the $L=36$ case (the largest size simulation), 
(iii) data as in (i) but without the $L=6$ case (the smallest size simulation). 
The corresponding error bars are computed with the resampling technique 
described in Methods of the main text. 
Notice that  for $U/t=4.4$ and $U/t=4.6$ only (ii) is available as, in these cases, we have not
performed  the largest size simulation.
The statistical errors are indicated by numbers 
in parentheses (corresponding to the last two digits). } 
\label{clmaxdt.4}
\end{table*}

\begin{table*}[htb]
\begin{tabular}{|l|c|c|c|c|c|} \hline
\backslashbox{$L$}{$U/t$} & 3.7 & 3.8 & 3.9 & 4   & 4.1\\ \hline
6               & 0.017641(50)      & 0.018946(98)      & 0.020350(77)      & 0.02147(13)       & 0.02332(13)      \\
9               & 0.007859(40)      & 0.008868(28)      & 0.009878(54)      & 0.011107(44)      & 0.012434(86)     \\
12              & 0.00473(21)       & 0.005204(15)      & 0.006078(24)      & 0.007090(90)      & 0.008227(39)     \\
15              & 0.002868(22)      & 0.003515(14)      & 0.004215(31)      & 0.005249(70)      & 0.006239(32)     \\
18              & 0.002043(18)      & 0.002571(17)      & 0.003210(28)      & 0.00404(04)       & 0.005011(27)     \\
24              & 0.001133(13)      & 0.0015855(67)     & 0.002183(21)      & 0.002904(36)      & 0.003890(27)     \\
36              & 0.000513(15)      & 0.0008547(53)     & 0.001271(19)      & 0.002014(72)      & 0.002943(31)     \\
$\infty$ (i)    & -0.000094(75)     & 0.000110(38)      & 0.000200(99)      & 0.00107(24)       & 0.00205(14)      \\
$\infty$ (ii)   & -0.00027(17)      & 0.00011(12)       & 0.00071(25)       & 0.00103(40)       & 0.00232(33)      \\
$\infty$ (iii)  & 0.00000(17)       & 0.000067(63)      & 0.00005(17)       & 0.00115(48)       & 0.00199(24)      \\
\hline \hline
\backslashbox{$L$}{$U/t$} & 4.2 & 4.3 & 4.4 & 4.5 & 4.6\\ \hline
6               & 0.02457(12)       & 0.02615(12)       & 0.02833(38)       & 0.03021(21)       & 0.03155(47)      \\
9               & 0.013687(49)      & 0.015186(58)      & 0.01706(32)       & 0.0187(02)        & 0.02022(24)      \\
12              & 0.009527(62)      & 0.010643(91)      & 0.012512(71)      & 0.014256(97)      & 0.0174(15)       \\
15              & 0.00763(15)       & 0.00874(12)       & 0.01056(16)       & 0.01181(15)       & 0.01380(60)      \\
18              & 0.006228(56)      & 0.00730(25)       & 0.00929(21)       & 0.01086(30)       & 0.0130(12)       \\
24              & 0.00519(11)       & 0.00573(10)       & 0.00837(52)       & 0.00881(22)       & 0.01069(60)      \\
36              & 0.00458(34)       & 0.00513(25)       &  ---              & 0.00827(59)       &  ---             \\
$\infty$ (i)    & 0.00372(63)       & 0.00373(64)       &  ---              & 0.0057(13)        &  ---             \\
$\infty$ (ii)   & 0.00336(73)       & 0.00286(81)       & 0.0083(27)        & 0.0046(15)        & 0.0038(55)       \\
$\infty$ (iii)  & 0.0053(12)        & 0.0038(14)        &  ---              & 0.0074(25)        &  ---             \\
\hline
\end{tabular}
\caption{Same as Table.~\ref{clmaxdt.4} but for $\Delta\tau t=0.2$.}
\label{clmaxdt.2}
\end{table*}

\begin{table*}[htb]
\begin{tabular}{|l|c|c|c|c|c|} \hline
\backslashbox{$L$}{$U/t$} & 3.7 & 3.8 & 3.9 & 4   & 4.1\\ \hline
6               & 0.01783(14)       & 0.018970(50)      & 0.020525(69)      & 0.02234(29)       & 0.02371(19)      \\
9               & 0.007859(20)      & 0.008791(19)      & 0.009842(30)      & 0.011013(33)      & 0.012464(72)     \\
12              & 0.004381(43)      & 0.005117(12)      & 0.005978(25)      & 0.006930(40)      & 0.008094(34)     \\
15              & 0.002796(18)      & 0.003446(20)      & 0.00412(03)       & 0.004999(52)      & 0.006109(41)     \\
18              & 0.001959(42)      & 0.002474(23)      & 0.003091(23)      & 0.003869(26)      & 0.004945(59)     \\
24              & 0.001096(11)      & 0.0015101(96)     & 0.002008(25)      & 0.002714(30)      & 0.003745(33)     \\
36              & 0.0004969(50)     & 0.0007721(79)     & 0.001181(20)      & 0.001901(81)      & 0.002727(42)     \\
$\infty$ (i)    & 0.000051(49)      & 0.000026(46)      & 0.000256(88)      & 0.00086(23)       & 0.00161(19)      \\
$\infty$ (ii)   & 0.00010(16)       & 0.00015(12)       & 0.00024(22)       & 0.00066(32)       & 0.00213(38)      \\
$\infty$ (iii)  & 0.00001(10)       & -0.000089(78)     & 0.00027(17)       & 0.00105(40)       & 0.00108(30)      \\
\hline \hline
\backslashbox{$L$}{$U/t$} & 4.2 & 4.3 & 4.4 & 4.5 & 4.6\\ \hline
6               & 0.025130(58)      & 0.02673(13)       & 0.02816(15)       & 0.03056(18)       & 0.03238(31)      \\
9               & 0.013821(61)      & 0.015357(81)      & 0.017086(95)      & 0.01945(47)       & 0.02106(73)      \\
12              & 0.00955(11)       & 0.01090(16)       & 0.012567(89)      & 0.01425(13)       & 0.01570(46)      \\
15              & 0.007286(49)      & 0.00855(11)       & 0.01029(15)       & 0.01171(16)       & 0.0134(02)       \\
18              & 0.005903(66)      & 0.00700(20)       & 0.00878(14)       & 0.01045(25)       & 0.0121(04)       \\
24              & 0.004874(76)      & 0.00608(15)       & 0.00765(21)       & 0.00913(27)       & 0.01085(64)      \\
36              & 0.00389(14)       & 0.00582(26)       &  ---              & 0.0103(22)        &  ---             \\
$\infty$ (i)    & 0.00318(40)       & 0.00594(74)       &  ---              & 0.0099(24)        &  ---             \\
$\infty$ (ii)   & 0.00319(58)       & 0.0046(11)        & 0.0054(13)        & 0.0095(23)        & 0.0116(43)       \\
$\infty$ (iii)  & 0.00426(83)       & 0.0087(15)        &  ---              & 0.0102(43)        &  ---             \\
\hline
\end{tabular}
\caption{Same as Table.~\ref{clmaxdt.4} but for $\Delta\tau t=0.1$.}
\label{clmaxdt.1}
\end{table*}

\begin{table*}[htb]
\begin{tabular}{|l|c|c|c|c|c|} \hline
\backslashbox{$L$}{$U/t$} & 3.7 & 3.8 & 3.9 & 4   & 4.1\\ \hline
6               & 0.01780(15)       & 0.019050(50)      & 0.020640(62)      & 0.02198(16)       & 0.02384(13)      \\
9               & 0.007851(21)      & 0.008794(20)      & 0.009833(32)      & 0.011032(35)      & 0.012506(65)     \\
12              & 0.004372(45)      & 0.005105(11)      & 0.005985(22)      & 0.006922(40)      & 0.008118(30)     \\
15              & 0.002789(16)      & 0.003424(15)      & 0.004067(32)      & 0.005019(49)      & 0.006108(32)     \\
18              & 0.001970(22)      & 0.002466(17)      & 0.003090(24)      & 0.003849(25)      & 0.004855(34)     \\
24              & 0.001070(11)      & 0.0014780(74)     & 0.002022(22)      & 0.002699(27)      & 0.003697(30)     \\
36              & 0.0004839(55)     & 0.0007488(57)     & 0.001147(21)      & 0.001879(66)      & 0.002719(32)     \\
$\infty$ (i)    & 0.000050(48)      & 0.000002(36)      & 0.000143(92)      & 0.00099(20)       & 0.00173(14)      \\
$\infty$ (ii)   & -0.00008(15)      & 0.00001(11)       & 0.00033(21)       & 0.00085(28)       & 0.00203(32)      \\
$\infty$ (iii)  & 0.00008(10)       & -0.000061(61)     & 0.00012(16)       & 0.00106(36)       & 0.00150(24)      \\
\hline \hline
\backslashbox{$L$}{$U/t$} & 4.2 & 4.3 & 4.4 & 4.5 & 4.6\\ \hline
6               & 0.025247(59)      & 0.02677(11)       & 0.02839(15)       & 0.03090(16)       & 0.03266(31)      \\
9               & 0.013819(49)      & 0.015395(62)      & 0.01716(10)       & 0.01909(22)       & 0.02082(30)      \\
12              & 0.009539(69)      & 0.01073(11)       & 0.012592(70)      & 0.01433(11)       & 0.01594(48)      \\
15              & 0.007283(51)      & 0.008566(96)      & 0.01038(14)       & 0.01180(14)       & 0.01348(20)      \\
18              & 0.005950(54)      & 0.00706(18)       & 0.00890(14)       & 0.01043(26)       & 0.01227(39)      \\
24              & 0.004866(75)      & 0.00577(15)       & 0.00764(23)       & 0.00879(28)       & 0.01094(57)      \\
36              & 0.00392(15)       & 0.00538(23)       &  ---              & 0.00880(79)       &  ---             \\
$\infty$ (i)    & 0.00307(39)       & 0.00482(65)       &  ---              & 0.0070(16)        &  ---             \\
$\infty$ (ii)   & 0.00291(55)       & 0.0034(10)        & 0.0054(12)        & 0.0056(18)        & 0.0104(36)       \\
$\infty$ (iii)  & 0.00435(73)       & 0.0060(13)        &  ---              & 0.0094(29)        &  ---             \\
\hline
\end{tabular}
\caption{Same as Table.~\ref{clmaxdt.4}, 
but $\Delta\tau$ is extrapolated to 0 by fitting the $\Delta\tau=0.4$, $0.2$, and $0.1$ data with 
quadratic polynomial in $\Delta\tau$ for each $L$. 
The extrapolations for $L\to\infty$ in (i), (ii), and (iii) are performed using the $\Delta\tau\to0$ 
extrapolated results for different $L$'s. }
\label{clmaxextr}
\end{table*}

\begin{table*}[htb]
\begin{tabular}{|l|c|c|c|c|c|} \hline
\backslashbox{$L$}{$U/t$} & 3.7 & 3.8 & 3.9 & 4   & 4.1\\ \hline
6               & 0.035808(97)      & 0.037194(23)      & 0.038750(51)      & 0.04014(11)       & 0.041717(46)     \\
9               & 0.017779(18)      & 0.018998(19)      & 0.020288(50)      & 0.021568(17)      & 0.0230043(91)    \\
12              & 0.010880(24)      & 0.0118617(61)     & 0.012927(21)      & 0.014172(13)      & 0.015485(15)     \\
15              & 0.007430(17)      & 0.008309(11)      & 0.009333(65)      & 0.010380(16)      & 0.011678(26)     \\
18              & 0.005495(18)      & 0.006260(12)      & 0.006932(41)      & 0.008199(12)      & 0.009405(14)     \\
24              & 0.003368(11)      & 0.0040547(51)     & 0.004737(21)      & 0.005831(17)      & 0.007026(12)     \\
36              & 0.0017280(89)     & 0.0023083(29)     & 0.002911(38)      & 0.003909(37)      & 0.005001(20)     \\
$\infty$ (i)    & -0.000260(50)     & 0.000200(24)      & 0.00078(16)       & 0.00142(11)       & 0.002433(68)     \\
$\infty$ (ii)   & -0.00046(14)      & 0.000121(82)      & 0.00070(27)       & 0.00137(15)       & 0.00259(12)      \\
$\infty$ (iii)  & -0.000156(99)     & 0.000207(39)      & 0.00099(25)       & 0.00161(17)       & 0.00239(12)      \\
\hline \hline
\backslashbox{$L$}{$U/t$} & 4.2 & 4.3 & 4.4 & 4.5 & 4.6\\ \hline
6               & 0.043429(70)      & 0.045119(28)      & 0.046740(33)      & 0.048433(35)      & 0.05051(61)      \\
9               & 0.024512(28)      & 0.026082(49)      & 0.027721(49)      & 0.029377(37)      & 0.030936(71)     \\
12              & 0.016901(23)      & 0.018318(75)      & 0.020054(39)      & 0.021689(46)      & 0.022623(73)     \\
15              & 0.012990(21)      & 0.014471(49)      & 0.016051(47)      & 0.01778(12)       & 0.01907(09)      \\
18              & 0.010766(37)      & 0.01202(10)       & 0.01404(24)       & 0.01519(19)       & 0.0170(02)       \\
24              & 0.008317(43)      & 0.009578(66)      & 0.01114(12)       & 0.01255(11)       & 0.01424(44)      \\
36              & 0.006268(77)      & 0.00781(25)       &  ---              & 0.01014(28)       &  ---             \\
$\infty$ (i)    & 0.00375(20)       & 0.00487(47)       &  ---              & 0.00630(54)       &  ---             \\
$\infty$ (ii)   & 0.00387(28)       & 0.00458(51)       & 0.00579(67)       & 0.00630(64)       & 0.0156(18)       \\
$\infty$ (iii)  & 0.00382(36)       & 0.0052(11)        &  ---              & 0.0061(14)        &  ---             \\
\hline
\end{tabular}
\caption{%
Spin structure factors $S_{\rm AF}/N$ for several values of $U/t$.  
Here $\Delta\tau t=0.4$ and projection time $\tau t=L+4$. 
The extrapolated values for $L\to\infty$ are obtained 
by cubic fits in $1/L$ using 
(i)  data corresponding to $L=6,9,12,15,18,24,36$, 
(ii) data as in (i) but without the $L=36$ case (the largest size simulation), 
(iii) data as in (i) but without the $L=6$ case (the smallest size simulation). 
The corresponding error bars are computed with the resampling technique 
described in Methods of the main text. 
Notice that  for $U/t=4.4$ and $U/t=4.6$ only (ii) is available as, in these cases, we have not
performed the largest size simulation.
The statistical errors are indicated by numbers 
in parentheses (corresponding to the last two digits). }
\label{spidt.4}
\end{table*}

\begin{table*}[htb]
\begin{tabular}{|l|c|c|c|c|c|} \hline
\backslashbox{$L$}{$U/t$} & 3.7 & 3.8 & 3.9 & 4   & 4.1\\ \hline
6               & 0.037612(28)      & 0.039255(45)      & 0.041014(29)      & 0.042784(74)      & 0.044697(46)     \\
9               & 0.018340(20)      & 0.019643(14)      & 0.020993(15)      & 0.022476(18)      & 0.024054(16)     \\
12              & 0.011030(23)      & 0.0120164(68)     & 0.013139(13)      & 0.014402(18)      & 0.015865(15)     \\
15              & 0.007408(14)      & 0.0082476(90)     & 0.009205(13)      & 0.010359(18)      & 0.0116936(87)    \\
18              & 0.0053791(90)     & 0.0060880(99)     & 0.006924(14)      & 0.00799(02)       & 0.009248(13)     \\
24              & 0.0031972(77)     & 0.0037857(29)     & 0.004508(11)      & 0.005467(20)      & 0.006654(10)     \\
36              & 0.0015324(71)     & 0.0019833(29)     & 0.002547(15)      & 0.003375(23)      & 0.004500(13)     \\
$\infty$ (i)    & -0.000373(36)     & -0.000092(19)     & 0.000152(58)      & 0.000761(83)      & 0.001724(51)     \\
$\infty$ (ii)   & -0.000570(95)     & -0.000157(64)     & 0.00018(11)       & 0.00089(17)       & 0.001736(97)     \\
$\infty$ (iii)  & -0.000268(76)     & -0.000085(30)     & 0.00016(10)       & 0.00065(15)       & 0.001810(97)     \\
\hline \hline
\backslashbox{$L$}{$U/t$} & 4.2 & 4.3 & 4.4 & 4.5 & 4.6\\ \hline
6               & 0.046595(44)      & 0.048480(74)      & 0.05085(15)       & 0.052760(89)      & 0.05521(25)      \\
9               & 0.025775(18)      & 0.027578(22)      & 0.029571(60)      & 0.03158(05)       & 0.033680(99)     \\
12              & 0.017409(16)      & 0.019056(51)      & 0.021001(26)      & 0.022981(59)      & 0.024851(98)     \\
15              & 0.013224(25)      & 0.014875(60)      & 0.016683(49)      & 0.018522(42)      & 0.02037(18)      \\
18              & 0.010672(17)      & 0.01222(16)       & 0.014087(51)      & 0.015852(79)      & 0.01789(30)      \\
24              & 0.008040(49)      & 0.009472(36)      & 0.011225(81)      & 0.013113(90)      & 0.01525(25)      \\
36              & 0.005907(41)      & 0.00735(12)       &  ---              & 0.01092(23)       &  ---             \\
$\infty$ (i)    & 0.00319(13)       & 0.00432(30)       &  ---              & 0.00781(48)       &  ---             \\
$\infty$ (ii)   & 0.00293(25)       & 0.00407(36)       & 0.00556(52)       & 0.00759(58)       & 0.0114(14)       \\
$\infty$ (iii)  & 0.00336(21)       & 0.00436(68)       &  ---              & 0.00874(95)       &  ---             \\
\hline
\end{tabular}
\caption{Same as Table.~\ref{spidt.4} but for $\Delta\tau t=0.2$.}
\label{spidt.2}
\end{table*}

\begin{table*}[htb]
\begin{tabular}{|l|c|c|c|c|c|} \hline
\backslashbox{$L$}{$U/t$} & 3.7 & 3.8 & 3.9 & 4   & 4.1\\ \hline
6               & 0.038120(65)      & 0.039777(32)      & 0.041607(45)      & 0.043467(73)      & 0.045326(52)     \\
9               & 0.018478(12)      & 0.0197460(90)     & 0.021142(16)      & 0.022633(14)      & 0.024266(15)     \\
12              & 0.010999(14)      & 0.0120133(85)     & 0.013127(11)      & 0.014399(15)      & 0.015846(15)     \\
15              & 0.007357(12)      & 0.0082117(80)     & 0.00914(01)       & 0.010291(18)      & 0.0116167(90)    \\
18              & 0.005295(11)      & 0.0060164(83)     & 0.006851(12)      & 0.007875(18)      & 0.009134(17)     \\
24              & 0.0031298(50)     & 0.0037011(36)     & 0.0043775(59)     & 0.005280(11)      & 0.0064862(86)    \\
36              & 0.0015024(20)     & 0.0018748(44)     & 0.0024468(99)     & 0.003225(28)      & 0.004250(23)     \\
$\infty$ (i)    & -0.000166(22)     & -0.000260(22)     & 0.000136(44)      & 0.000568(86)      & 0.001478(69)     \\
$\infty$ (ii)   & -0.000349(81)     & -0.000293(54)     & 0.000004(85)      & 0.00045(12)       & 0.001626(91)     \\
$\infty$ (iii)  & -0.000104(44)     & -0.000258(38)     & 0.000182(71)      & 0.00063(16)       & 0.00121(13)      \\
\hline \hline
\backslashbox{$L$}{$U/t$} & 4.2 & 4.3 & 4.4 & 4.5 & 4.6\\ \hline
6               & 0.047429(27)      & 0.049415(49)      & 0.051481(68)      & 0.05401(24)       & 0.05610(12)      \\
9               & 0.026021(30)      & 0.027858(19)      & 0.029900(36)      & 0.031952(33)      & 0.03438(27)      \\
12              & 0.017458(35)      & 0.019281(89)      & 0.021168(35)      & 0.023206(63)      & 0.02508(17)      \\
15              & 0.013083(30)      & 0.014722(39)      & 0.016661(89)      & 0.018642(67)      & 0.02070(08)      \\
18              & 0.010558(32)      & 0.01212(11)       & 0.014063(72)      & 0.015833(76)      & 0.0179(01)       \\
24              & 0.007796(26)      & 0.009361(42)      & 0.01113(12)       & 0.01314(23)       & 0.01498(15)      \\
36              & 0.005492(66)      & 0.00752(13)       &  ---              & 0.01022(22)       &  ---             \\
$\infty$ (i)    & 0.00252(18)       & 0.00466(31)       &  ---              & 0.00609(60)       &  ---             \\
$\infty$ (ii)   & 0.00260(24)       & 0.00413(37)       & 0.00534(60)       & 0.0060(10)        & 0.0094(13)       \\
$\infty$ (iii)  & 0.00246(36)       & 0.00685(78)       &  ---              & 0.0068(10)        &  ---             \\
\hline
\end{tabular}
\caption{Same as Table.~\ref{spidt.4} but for $\Delta\tau t=0.1$.}
\label{spidt.1}
\end{table*}

\begin{table*}[htb]
\begin{tabular}{|l|c|c|c|c|c|} \hline
\backslashbox{$L$}{$U/t$} & 3.7 & 3.8 & 3.9 & 4   & 4.1\\ \hline
6               & 0.038233(42)      & 0.039947(30)      & 0.041782(33)      & 0.043679(65)      & 0.045621(43)     \\
9               & 0.018525(12)      & 0.0198080(89)     & 0.021207(16)      & 0.022723(13)      & 0.024369(13)     \\
12              & 0.011020(14)      & 0.0120453(66)     & 0.013158(10)      & 0.014433(14)      & 0.015918(13)     \\
15              & 0.007368(11)      & 0.0082122(73)     & 0.009129(13)      & 0.010311(15)      & 0.0116376(86)    \\
18              & 0.0053070(91)     & 0.0060089(77)     & 0.006863(12)      & 0.007879(15)      & 0.009158(13)     \\
24              & 0.0031187(50)     & 0.0036857(29)     & 0.0043622(63)     & 0.005259(11)      & 0.0064725(80)    \\
36              & 0.0014867(23)     & 0.0018627(31)     & 0.002417(10)      & 0.003187(23)      & 0.004282(15)     \\
$\infty$ (i)    & -0.000206(21)     & -0.000239(18)     & 0.000075(45)      & 0.000497(74)      & 0.001493(53)     \\
$\infty$ (ii)   & -0.000449(72)     & -0.000315(49)     & -0.000035(85)     & 0.00036(11)       & 0.001506(82)     \\
$\infty$ (iii)  & -0.000121(44)     & -0.000207(30)     & 0.000105(74)      & 0.00051(13)       & 0.001494(99)     \\
\hline \hline
\backslashbox{$L$}{$U/t$} & 4.2 & 4.3 & 4.4 & 4.5 & 4.6\\ \hline
6               & 0.047688(28)      & 0.049679(47)      & 0.051846(68)      & 0.05424(11)       & 0.05649(14)      \\
9               & 0.026167(21)      & 0.028002(18)      & 0.030071(36)      & 0.032170(31)      & 0.03460(12)      \\
12              & 0.017556(20)      & 0.019318(59)      & 0.021280(27)      & 0.023350(53)      & 0.02547(11)      \\
15              & 0.013186(24)      & 0.014794(38)      & 0.016830(56)      & 0.018737(52)      & 0.020809(84)     \\
18              & 0.010605(21)      & 0.01216(10)       & 0.014077(69)      & 0.015934(72)      & 0.01800(15)      \\
24              & 0.007789(26)      & 0.009384(35)      & 0.011197(88)      & 0.01328(11)       & 0.01511(16)      \\
36              & 0.005642(47)      & 0.00738(12)       &  ---              & 0.01057(20)       &  ---             \\
$\infty$ (i)    & 0.00262(14)       & 0.00442(27)       &  ---              & 0.00716(47)       &  ---             \\
$\infty$ (ii)   & 0.00225(20)       & 0.00404(32)       & 0.00525(48)       & 0.00733(65)       & 0.0098(11)       \\
$\infty$ (iii)  & 0.00295(24)       & 0.00588(64)       &  ---              & 0.00806(87)       &  ---             \\
\hline
\end{tabular}
\caption{Same as Table.~\ref{spidt.4},
but $\Delta\tau$ is extrapolated to 0 by fitting the $\Delta\tau=0.4$, $0.2$, and $0.1$ data with 
quadratic polynomial in $\Delta\tau$ for each $L$. 
The extrapolations for $L\to\infty$ in (i), (ii), and (iii) are performed using the $\Delta\tau\to0$ 
extrapolated results for different $L$'s. }
\label{spiextr}
\end{table*}

Finally, the spin gap $\Delta_s$ is estimated by directly calculating the ground state energies for 
the spin singlet sector [$E(S=0)$] and for the spin triplet sector [$E(S=1)$] separately, i.e., 
\begin{equation}
\Delta_s = E(S=1) - E(S=0). 
\end{equation}  
The results are shown in TABLE~\ref{sgap} calculated for various system sizes with $L$ up to $18$ and for 
different values of $U/t$. The thermodynamically extrapolated values for $L\to\infty$ are estimated by fitting the results 
for different $L$'s with quadratic polynomials in $1/L$ and 
these results are also shown in the last row of TABLE~\ref{sgap}. 

\begin{table*}[htb]
\begin{tabular}{|l|c|c|c|c|}
\hline
 \backslashbox{$L$}{$U/t$}  & 3.4 & 3.7 & 4  & 4.3  \\
\hline
6          & 0.11452(62)       & 0.13034(48)       & 0.14054(68)       & 0.14287(94)      \\
9          & 0.0828(11)        & 0.09672(72)       & 0.1032(10)        & 0.1020(22)       \\
12         & 0.06292(97)       & 0.07611(99)       & 0.08212(92)       & 0.0753(24)       \\
15         & 0.0524(11)        & 0.0634(12)        & 0.06743(92)       & 0.0616(26)       \\
18         &  ---              &  ---              & 0.0558(10)        &  ---             \\
$\infty$   & -0.0009(71)       & 0.0021(61)        & -0.0018(42)       & -0.013(16)       \\
\hline
\end{tabular}
\caption{ The spin gap $\Delta_S= E(S=1)-E(S=0)$ for several values of $U/t$. 
Here we set $\Delta\tau t=0.14$ and projection time $\tau t=L+4$. 
The extrapolated values for $L\to\infty$ are obtained  by a quadratic 
fit in $1/L$ and the corresponding error bars are computed with the   
resampling technique described in Methods of the main text. The statistical errors are indicated by numbers 
in parentheses (corresponding to the last two digits). 
} 
\label{sgap}
\end{table*}

\end{document}